\documentclass[preprint,12pt]{elsarticle}

\usepackage{amssymb}

\newcommand{\tref}[1]{(\ref{#1})}

\newcommand{\tprecomment}[1]{ #1 } 

\newcommand{\bea}{\begin{eqnarray}}
\newcommand{\eea}{\end{eqnarray}}
\newcommand{\beq}{\begin{equation}}
\newcommand{\eeq}{\end{equation}}

\newcommand{\ks}{k_\mathrm{s}}
\newcommand{\kav}{\langle k \rangle}
\newcommand{\ksav}{\langle \ks \rangle}

\journal{Physica A}

\begin{document}

\begin{frontmatter}




\title{The Emergence of Leadership in Social Networks\tnoteref{preprintnumber}}
\tnotetext[preprintnumber]{Accepted for publication in Physica A. \texttt{Imperial/TP/11/TSE/3}, \texttt{arXiv:1106.0296}.}


\author[ImperialCollege]{T.\ Clemson}

\author[ImperialCollege]{T.S.\ Evans}
\ead[http://www.imperial.ac.uk/people/T.Evans]{http://www.imperial.ac.uk/people/T.Evans}

\address[ImperialCollege]{Theoretical Physics, Physics Department,
Imperial College London, South Kensington campus, London, SW7 2AZ, UK}

\begin{abstract}
We study a networked version of the minority game in which agents can choose to follow the choices made by a neighbouring agent in a social network.  We show that for a wide variety of networks a leadership structure always emerges, with most agents following the choice made by a few agents.  We find a suitable parameterisation which highlights the universal aspects of the behaviour and which also indicates where results depend on the type of social network.
\end{abstract}



\end{frontmatter}


\section{Introduction}
    \label{cha:introduction}
        The minority game is a simple model for competition dynamics that attempts to explain the {behaviour} of a collection of {agents} competing for a {limited resource} \citep{CZ97,M04,CMZ05}. Traditionally, such games have been used to provide a simple platform for experimentation in financial market dynamics but their applicability is far more general. Any situation in which it is beneficial to be in the minority or to possess some information that few others possess will exhibit characteristics inherent in the minority game.

        In recent years, the study of large and complicated networks, such as vast {social networks} or intricate {genetic networks}, has seen a huge rise in popularity \citep{BA99,Evans04,BBV08}. Partly this is due to the availability of data but also because these networks often constrain the dynamics of interactions from which new phenomena emerge.

It is then natural to ask how the dynamics of the minority game
are modified by the presence of a substrate network which
represents the `social' network linking the agents.  Previous
studies have combined the minority game with regular networks, in
one \cite{KSB00,S00a,S01a,CCH04,CD06,LS07a} or two \cite{MD02,CC04,K06,CD06}
dimensions, and with random networks of various types:
Erd\H{o}s-R\'{e}nyi (or classical) random graphs
\cite{ATBK04,LCHJ04,CJH04,LCHJ05a,LJ06c,CD06,LS07a}, Kaufman Boolean random
networks \cite{PBC00,GL02} and scale-free networks
\cite{LJ06c,CD06,LS07a}.  In all cases the networks are providing a
local source of information to the agents which usually
supplements the global information available, namely which choice
formed the minority in previous time steps. The local information
can be implemented in many different ways but here we will just
allow agents to copy the choices made by their nearest neighbours
in the substrate network as used in \cite{S00a,ATBK04}.

In particular an earlier study by Anghel et al.\ \cite{ATBK04}
noted that the probability that an agent is followed by $k$ other
agents is roughly proportional to $k^{-1}$ when the substrate
network is an Erd\H{o}s-R\'{e}nyi random graph.  That is, despite
the fact that there is no power law in the degree distribution of
the agent's social network, a few 'leaders' naturally emerge from
amongst the agents, 90\% of whom are merely `followers'.  It has
been suggested that this may simply be due to the underlying
copying processes, which in the simpler Moran model often leads to
a power law of the form $k^{-\gamma}$ with $\gamma \leq 1$
\cite{E04,E07,EP07,EPY10}. It is interesting to note that purely
local processes can give rise to power law behaviour in the degree
distribution of growing networks \cite{ES05,SLOJ08} so such
behaviour in a networked minority game is not so surprising.

In this paper we shall investigate two aspects of the leadership
structure found in the networked minority game.  First we shall
see if the power-law leadership structure of \cite{ATBK04} is robust when we change
the substrate network and secondly, we shall search for a
universal characterisation of the emergent leadership structure
and compare this with that found in the Moran model.

\section{Model}
        \label{sec:background}

\subsection{The Standard Minority Game}
            \label{sub:the_standard_minority_game}

The standard minority game is played by an odd number, $N$, of {agents}. A {time step} in the game consists of each agent trying to predict which of two abstract {outcomes}, $A$ or $B$, the fewest of its peers will also predict. Once all agents have chosen, the outcome picked by the fewest agents is deduced and each of the agents in this {minority} group is rewarded with a point for winning. The game continues in this way for an arbitrary number of time steps. 

                \begin{figure*}[bthp!]
                    \centering
\includegraphics[width=0.4\textwidth]{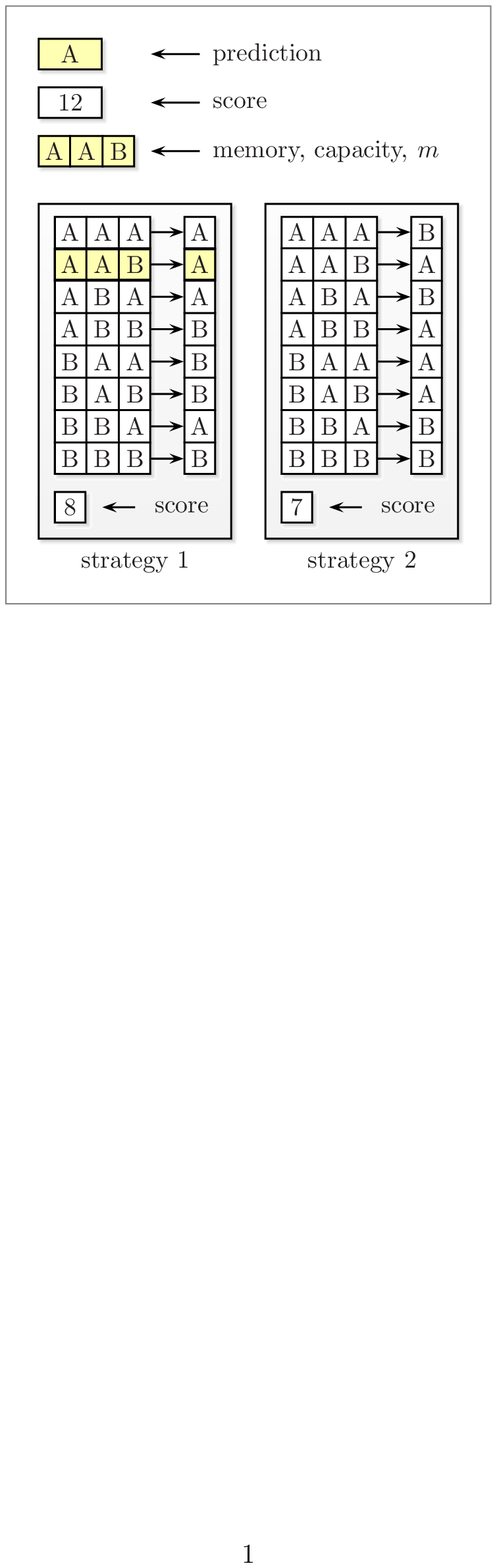}
                    \caption{
                        The potential contents of an agent's brain at an arbitrary time, for the case where $m=3$, $S=2$, including the agent's memory, score and strategies. The highlighted mapping represents that which would be used if this agent was asked to make a prediction for the time step. 
                    }
                    \label{fig:agent-brain}
                \end{figure*}

In order to make a prediction of the next minority outcome, each agent is equipped with a {brain}, consisting of a {memory} with a finite capacity, $m$, of past minority outcomes and a set of $S$ {strategies} mapping possible memory contents to predictions. Each strategy in the set has an associated score which represents the number of times that strategy would have predicted correctly in the past time steps. At the start of the game, the agent memories are each initialised to the same random string of past minority outcomes. When an agent must make a prediction, it selects the best performing strategy, based on score, from the set and uses the outcome prediction mapped by the current contents of its memory. If more than one strategy has the same score, the strategy used is chosen randomly from this subset with uniform probability. At the end of each time step, each agent updates the contents of its brain to reflect the outcome of the time step, by incrementing its score if it predicted correctly and updating the contents of its memory. Figure~\ref{fig:agent-brain} depicts the possible contents of an agent's brain at an arbitrary time step.

As Figure~\ref{fig:agent-brain} demonstrates, for a memory
capacity of $m=3$, each strategy contains $2^{3}=8$ mappings. In
general, for a memory capacity, $m$, there are $2^{m}$ possible
permutations of $A$ and $B$ and thus $2^{m}$ mappings in the
strategy.  Since each possible memory string can independently map
to either $A$ or $B$, the total number of distinct strategies for
a memory capacity, $m$, is $2^{2^{m}}$ \citep{CMZ05}. The set of
all possible strategies for a memory capacity, $m$ is called the
{strategy space}. At the start of the game, each agent is provided
with $S$ strategies at random from this strategy space. The size
of the strategy space relative to the number of agents in the game
is important in characterising the game dynamics.

            \subsubsection{The Networked Minority Game}
            \label{sub:the_networked_minority_game}

                In the standard minority game, each agent is equipped with access to the same {global information} as all of its peers in the form of a fixed capacity memory of past minority outcomes. In this way, agent behaviour (`intelligence') can only be varied by varying the memory capacity or by changing number of strategies an agent has for acting on this memory.

                The networked minority game aims to provide each agent with further {local information} in terms of access to a subset of the other agent's predictions. This effectively provides each agent with a {neighbourhood} with which it can {communicate}, providing a much broader range of agent intelligence in the game. In order to achieve this, each agent is considered as a vertex in a {social network} with the edges between them representing {communication channels}. We will refer to this network as the \emph{substrate network}. In our work this is taken to be a simple graph (no edge weights or self-loops) and we denote the degree of vertices in this network as $\ks$.

                In this regime, a time step initially progresses in a similar way to the standard minority game in that an agent predicts the minority outcome for that time step using its personal strategies and the global past minority outcome information. Rather than using this prediction as the final decision for the time step, the agent publicises it to each of its friends in its neighbourhood of the substrate network.

                Once all agents have publicised their predictions, the agent can make its final decision by following the prediction of the friend that has predicted correctly most frequently in past time steps. Thus, each agent must also keep track of how often its predictions were correct. If more than one friend scores equally and highest of all friends in terms of correct predictions, one friend is chosen from this subset at random with uniform probability. If the agent itself has predicted correctly most frequently, it follows its own prediction.

                \begin{figure*}[bthp!]
                    \centering
                    \includegraphics[width=0.8\textwidth]{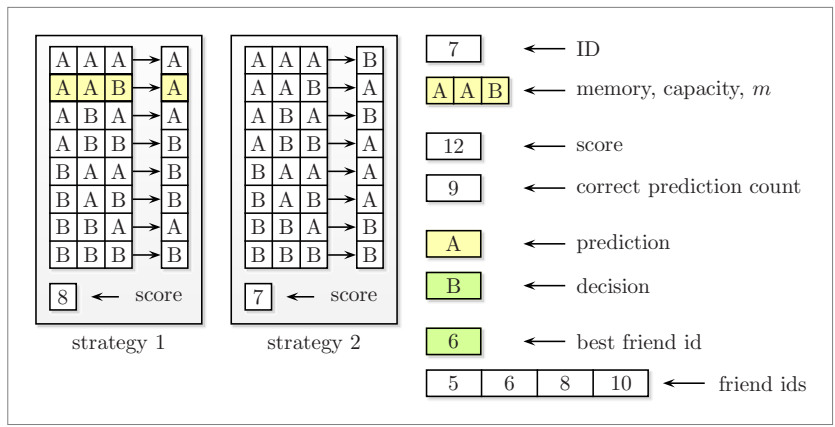}
                    \caption{
                        The potential contents of a networked agent's brain for $m=3$, $S=2$. In comparison to Figure~\ref{fig:agent-brain}, the networked agent's brain maintains an ID, a correct prediction count and has separate stores for the current prediction and decision. The agent's brain also has access to a list of the IDs of the agent's friends and maintains a record of the most recently followed or {best} friend's ID. This is used in constructing the influence network. The yellow highlighting indicates the source of the prediction whilst the green highlighting indicates the source of the decision. 
                    }\label{fig:networked-agent-brain}
                \end{figure*}

Note, it is important to differentiate between a {prediction} and a final {decision}. The convention adopted here is that a prediction is deduced from an agent's strategies whereas a final decision is deduced from the predictions of nearest neighbour friends. Separate scores are maintained for the number of times an agent's prediction has been correct and the number of times an agent's final decision has been correct. Our procedure is identical to that of \cite{ATBK04} but we note that allowing agents to follow strategies of non-nearest neighbours (they copy the decisions not the predictions of nearest neighbours) is an alternative approach used in \cite{CD06,LS07}.

Figure~\ref{fig:networked-agent-brain} shows how the brain of an agent in the networked game differs to that of an agent in the standard game. 

\begin{figure*}[tbhp!]
                \centering
\includegraphics[width=0.95\textwidth]{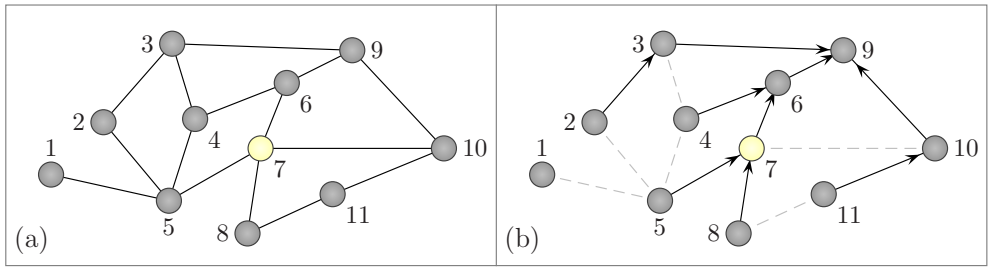}
\caption{
                    (a) An example substrate network for the case where $N=11$. (b) A possible influence network at some arbitrary time for the substrate network given in (a); an arrow from agent $i$ to agent $j$ means agent $i$ is following agent $j$'s prediction for this time-step; a disconnected agent such as agent $1$ is following its own prediction. The highlighted vertex, numbered 7, corresponds to the agent with brain depicted in Figure~\ref{fig:networked-agent-brain} and has degrees $\ks=3$ from the substrate network and $k=2$ from the influence network.
                }
                \label{fig:substrate-and-influence-networks}
            \end{figure*}

The decisions made then define a second network, which we will call the \emph{influence network}.  This is a directed subgraph of the substrate network in which the directed edges represent one agent following the prediction of another. The out-degree of vertices is always one.  It is the in-degree of vertices in this influence network which indicates the number of followers each agent has.  We will simply denote this degree as $k$ with pdf $p(k)$ (and $n(k)=Np(k)$).  Each agent has two non-trivial degrees, one from each network with $k<\ks$, and Figure~\ref{fig:substrate-and-influence-networks} demonstrates the difference between these networks through an example.

Observables studied in the standard minority game, such as choice attendance, variance and wealth (see for example \cite{CM03,CZ97,SMR99}), are just as relevant in the networked minority game.  Here, however, we will focus only on features relevant in the presence of a network. In particular, following Anghel et al.\ \cite{ATBK04}\footnote{Some related work has also been conducted by others, for instance, see \citep{LS07}.} we will look for a power law in the degree distribution of the influence network  as this hints at some self-organisation amongst the agents in the game.

    \section{Method}
    \label{cha:method}

\subsection{Measurement Requirements}
        \label{sec:ensemble_and_time_averages}

Since the initial contents of the agent memories is chosen at random, it is necessary to allow the game to run for some number of {settling} time steps so that it stabilises. Here a settling time of $100\times2^{m}$ was used since this value was found to be sufficient in previous work \citep{SMR99}.

In addition to the disorder in the memory contents, there is also disorder present in the strategies in use in the game since each execution may involve different strategies from the strategy space. Similarly when generating influence network data sets, since the substrate networks each involve an element of randomness, it is possible that the equilibrium reached will vary. Thus it is important to average over these differences to find the equilibrium state of the game in each case. Two different averages were used throughout: a {time} average where quantities were averaged over time steps and an {ensemble} average where quantities were averaged over distinct executions of the game with the same parameters.

Separately, the influence network is a function of time and so to find the correct influence network degree distribution, an average must be taken over many time steps. Since not all agents will change their leader, there is likely to be correlation in the influence network between subsequent time steps. It was deduced, through a comparison with past results, that a time difference between influence network measurements of $100$ time steps was sufficient.

\subsection{Generation of Influence Network Degree Distributions}
        \label{sec:generation_of_influence_network_degree_distributions}

The most important measurable in later sections is the in-degree distribution of the influence network, $p(k)$, formed between the agents in the game. The following details the way in which the model is used to generate this influence network degree distribution:

            \begin{enumerate}
                \item Construct a game using a substrate network of the required type and with the required parameters,
                \item Perform a number of time steps equal to the required settling time,
                \item For the number of influence networks over which the degree distribution should be time averaged:
                \begin{enumerate}
                    \item Generate the influence network using the best friend information stored by each agent,
                    \item Store the degree distribution for that influence network,
                    \item Iterate a number of time steps equal to the required influence network measurement spacing,
                \end{enumerate}
                \item Find and store the time average of the influence network degree distribution,
                \item Repeat for as many runs required for the ensemble average.
            \end{enumerate}

\subsection{Graph Algorithms}
        \label{sec:graph_algorithms}

Three different substrate networks were used: Erd\H{o}s-R\'{e}nyi random
graph (random or ER for short), a scale-free network (SF)
and a regular-ring (Ring) network representing a wide range of different degree distributions.

The random graph substrate was an Erd\H{o}s-R\'{e}nyi random
graph.  Here all $N(N-1)/2$ agent pairs are considered and an edge
is added with probability $p$ where $\ksav = p(N-1)$ is the
desired average degree in the substrate network.  The scale-free
graph is created using the JUNG library generator \cite{jung}
which generates networks with a $\ks^{-3}$ power law degree
distribution using the preferential attachment algorithm of
\citep{BA99}. Finally for the regular ring network, the $N$
vertices are placed on a ring and then each vertex is connected to
the next $\ksav/2$ vertices in each direction around the ring.

    \section{Results and Observations}
    \label{cha:results_and_observations}
Our computational model was checked in two ways. Firstly we
reproduced the standard minority game for the normalised variance
in choice attendance, $\frac{\sigma^{2}}{N}$, as a function of
memory capacity, $m$, number of agents, $N$, and number of
strategies per agent, $S$ \citep{CMZ05}.
\tprecomment{Figure~\ref{fig:standard-minority-game-verification}
given in the appendix shows our plot.}  In light of these results,
we decided to use two strategies, $S=2$, and a memory capacity of
six, $m=6$, so as to produce a low normalised variance in choice
attendance in the standard minority game.

The presence of the substrate network will alter the relationship between $m$, $S$ and the behaviour of the Minority
game, e.g.\ the normalised variance in choice attendance \cite{LS07a}.  As we are interested in studying and extending the results of \cite{ATBK04} on leadership structures, and given the majority of results in \cite{ATBK04} are for $S=2$ and $m=6$, we only work with these values. Our focus is on the variation of the substrate network parameters.

\begin{figure*}[btph!]
                    \centering
                    \includegraphics[width=0.8\textwidth]{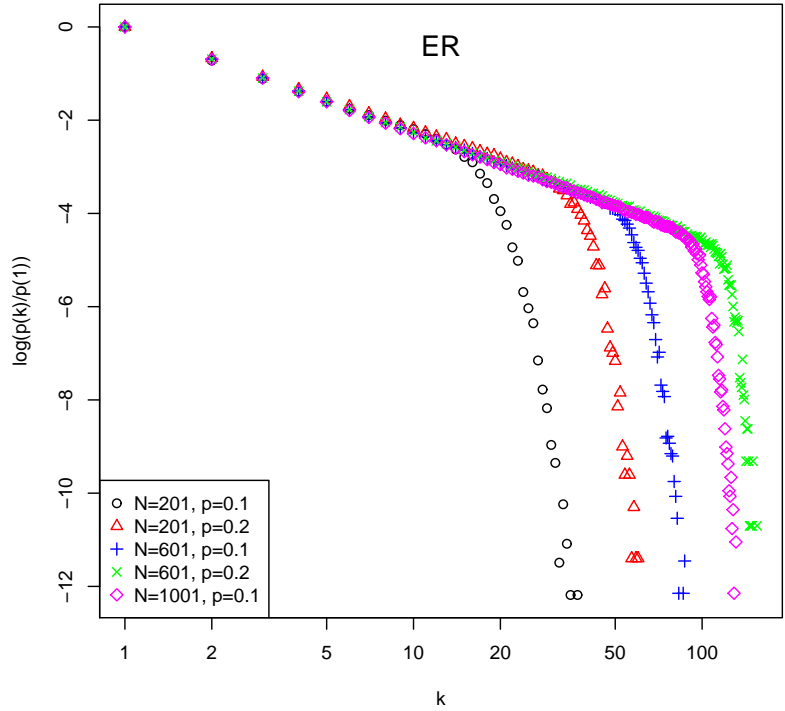}
                    \caption{Influence network degree distribution for a random substrate network as a function of $N$ and $p$. Each data set was generated for $m=6$, $S=2$ with a settling time of $100000$ time steps, time averaged over $1000$ influence networks each spaced by $100$ time steps and ensemble averaged over $20$ realisations of the game. The number of agents with zero followers has been excluded from the graph and the distribution has been normalised by the number of agents with one follower. The standard error in each data point is in all cases smaller than the data point itself.}
                    \label{f:ERpkp1}
                \end{figure*}

In a similar way, the networked minority game was verified by
generating data to plot the influence network degree distribution
given in Figure~1a of \citep{ATBK04}. Our results, shown in
Figure~\ref{f:ERpkp1}, compare very well.

                As can be seen in Figure~\ref{f:ERpkp1}, the power-law nature of the influence network has been confirmed. This occurs despite the substrate network being random. \citet{ATBK04} deduced that this occurs because those agents that predict correctly most often are more likely to be followed, generating a power-law influence network in a way similar to the Moran model \cite{E04,E07,EP07,EPY10}.

                What still remains unexplained is the cutoff point in the degree distribution after which
                it decays very quickly. Further, the emergence of a scale-free influence network has only been confirmed for a random substrate network. In an attempt to provide a more complete understanding of the influence network structure, the following sections perform further analysis on the influence network degree distribution for a random substrate network as well as investigating the effect of using various different substrate networks.

\begin{table}[htbp]
\begin{tabular}{|c|c|c||l|l|l|l|l|}
\hline
\multicolumn{3}{|c|}{\parbox{2.5cm}{ER Random \\ substrate}} &
                                \multicolumn{5}{|c|}{
                                    $   p(k)=A(k+k_{1})^{-\gamma}(1+\frac{k}{k_{2}})^{-\chi}  $
                                } \\ \hline
$N$ & $\ksav$ & $p$ & $k_1$ & $k_2$ & $\chi$ & $\gamma$ & $\ln(a)$ \\ \hline
 201  &  20.1 & 0.10 & 0.5$\pm$1.2 & 20.9$\pm$0.5 & 15.0$\pm$0.6 & 1.2$\pm$0.3 & 2.3$\pm$0.3 \\ \hline
 201  &  39.8 & 0.20 & 0.5$\pm$0.6 & 41.5$\pm$0.4 & 19.7$\pm$0.6 & 1.1$\pm$0.1 & 1.5$\pm$0.2 \\ \hline
 601  &  60.1 & 0.10 & 0.5$\pm$0.5 & 61.5$\pm$0.4 & 21.1$\pm$0.5 & 1.08$\pm$0.07 & 2.1$\pm$0.2 \\ \hline
 601  & 120.0 & 0.20 & 0.5$\pm$0.4 & 125.1$\pm$0.5 & 29.7$\pm$0.9 & 1.05$\pm$0.04 & 1.4$\pm$0.2 \\ \hline
 1001 & 100.1 & 0.10 & 0.4$\pm$0.3 & 103.2$\pm$0.4 & 26.4$\pm$0.5 & 1.05$\pm$0.03 & 2.1$\pm$0.1 \\ \hline \hline
\multicolumn{3}{|c|}{\parbox{2.5cm}{Scale Free \\ substrate}} &
                                \multicolumn{5}{|c|}{
                                    $   p(k)=A(k+k_{1})^{-\gamma}(1+\frac{k}{k_{2}})^{-\chi}  $
                                } \\ \hline
$N$ & $\ksav$ & $p$ & $k_1$ & $k_2$ & $\chi$ & $\gamma$ & $\ln(a)$ \\ \hline
201  &  20.8 & 0.10 & 0.5$\pm$1.4 & 14.8$\pm$0.7 & 6.5$\pm$0.3 & 1.1$\pm$0.4 & 2.5$\pm$0.3 \\ \hline
201  &  42.3 & 0.21 & 0.5$\pm$1.0 & 34.4$\pm$1.4 & 6.3$\pm$0.2 & 1.1$\pm$0.2 & 1.6$\pm$0.3 \\ \hline
401  &  86.0 & 0.22 & -0.4$\pm$0.4 & 60.8$\pm$1.7 & 5.8$\pm$0.1 & 0.87$\pm$0.08 & 1.8$\pm$0.3 \\ \hline
601  &  62.4 & 0.10 & -0.3$\pm$0.5 & 40.6$\pm$1.3 & 6.0$\pm$0.1 & 0.87$\pm$0.12 & 2.6$\pm$0.3 \\ \hline
601  & 129.8 & 0.22 & -0.5$\pm$0.3 & 91.0$\pm$2.0 & 5.7$\pm$0.1 & 0.87$\pm$0.06 & 1.8$\pm$0.3 \\ \hline
1001 & 104.0 & 0.10 & -0.4$\pm$0.3 & 68.0$\pm$1.4 & 5.8$\pm$0.1 & 0.89$\pm$0.07 & 2.6$\pm$0.3 \\ \hline \hline
\multicolumn{3}{|c|}{\parbox{2.5cm}{Ring \\ substrate}} &
                                \multicolumn{5}{|c|}{
                                    $   p(k)=A(k+k_{1})^{-\gamma}\exp\{-(\frac{k}{k_{2}})^{\chi}\} $
                                } \\ \hline
$N$ & $\ksav$ & $p$ & $k_1$ & $k_2$ & $\chi$ & $\gamma$ & $\ln(A)$ \\ \hline
201 & 20 & 0.1 & 0.5$\pm$1.1 & 14.0$\pm$1.3 & 2.4$\pm$0.3 & 0.5$\pm$0.2 & 1.7$\pm$0.4 \\ \hline
201 & 40 & 0.2 & 0.5$\pm$0.8 & 26.0$\pm$1.5 & 2.2$\pm$0.2 & 0.43$\pm$0.11 & 0.6$\pm$0.2 \\ \hline
401 & 40 & 0.1 & 0.5$\pm$0.7 & 47.7$\pm$2.2 & 2.0$\pm$0.1 & 0.35$\pm$0.06 & 0.1$\pm$0.2 \\ \hline
601 & 60 & 0.1 & 0.5$\pm$0.8 & 39.4$\pm$2.0 & 2.1$\pm$0.2 & 0.36$\pm$0.08 & 0.8$\pm$0.2 \\ \hline
601 & 120 & 0.2 & 0.5$\pm$0.6 & 75.7$\pm$2.4 & 2.1$\pm$0.1 & 0.36$\pm$0.04 & -0.3$\pm$0.1 \\ \hline
\end{tabular}
\caption{Fit functions and parameters with standard errors for the influence network degree distributions arising from random, scale-free and regular-ring substrate networks. Note that for double power law the normalisation $A$ is given in terms of $a=A(1+k_{1})^{-\gamma}(1+\frac{1}{k_{2}})^{-\chi}$.}
\label{tab:fit-parameters}
\end{table}

                \subsection{Random Substrate Network}
                \label{ssub:random_substrate_network}
\begin{figure*}[ptbh!]
                    \begin{center}
                        \includegraphics[width=0.49\textwidth]{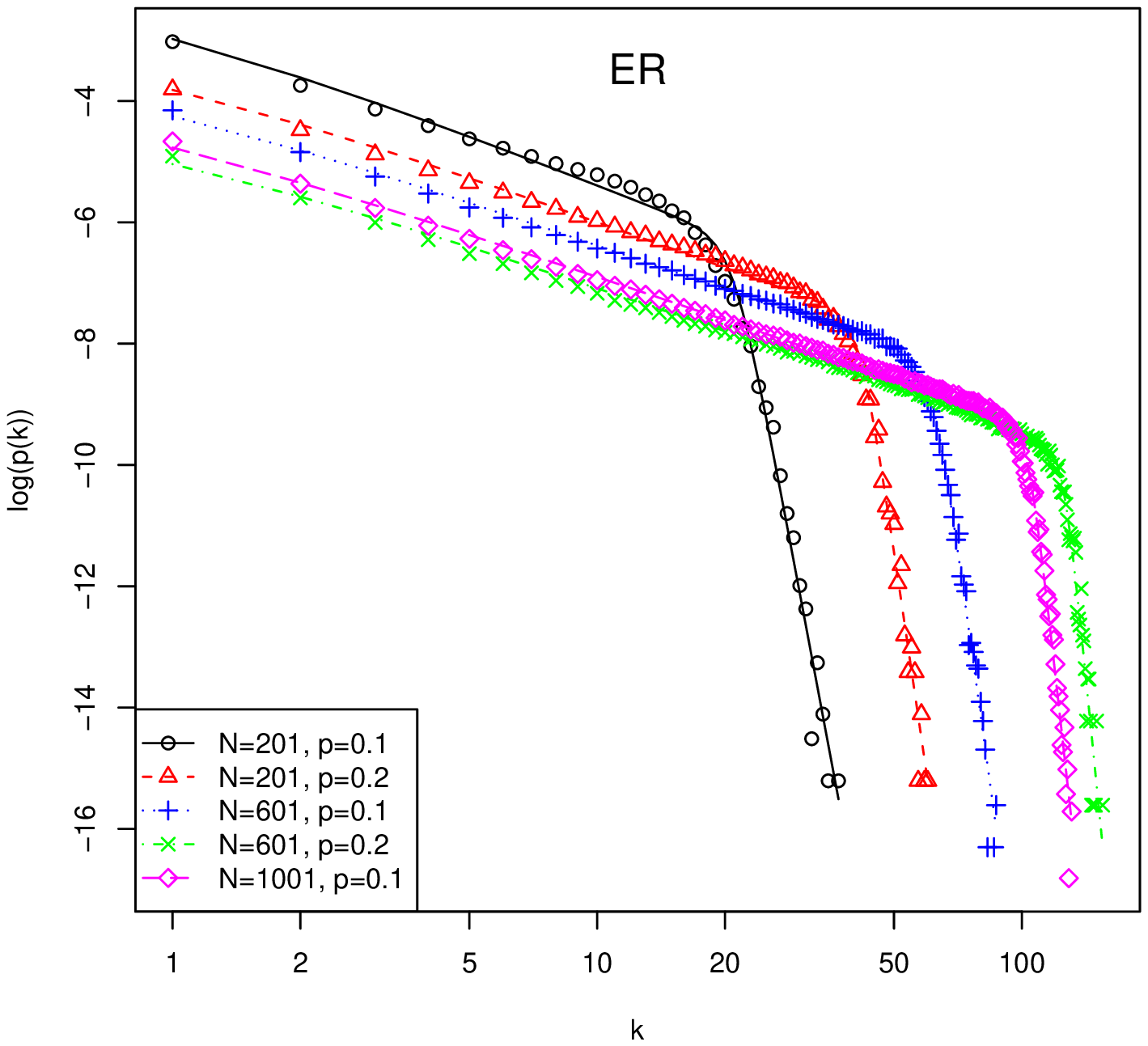}
                        \includegraphics[width=0.49\textwidth]{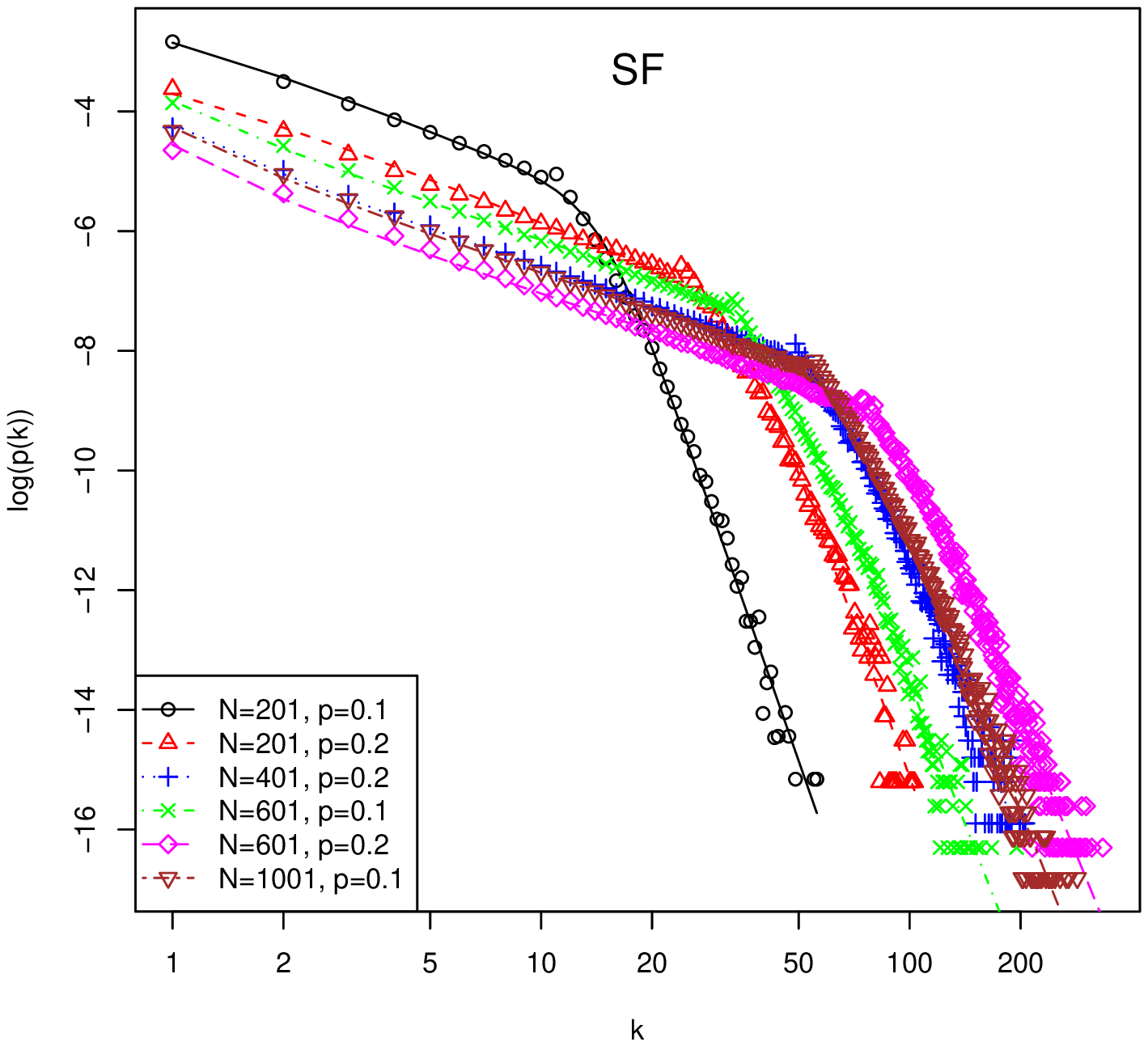}
                        \\
                        \includegraphics[width=0.49\textwidth]{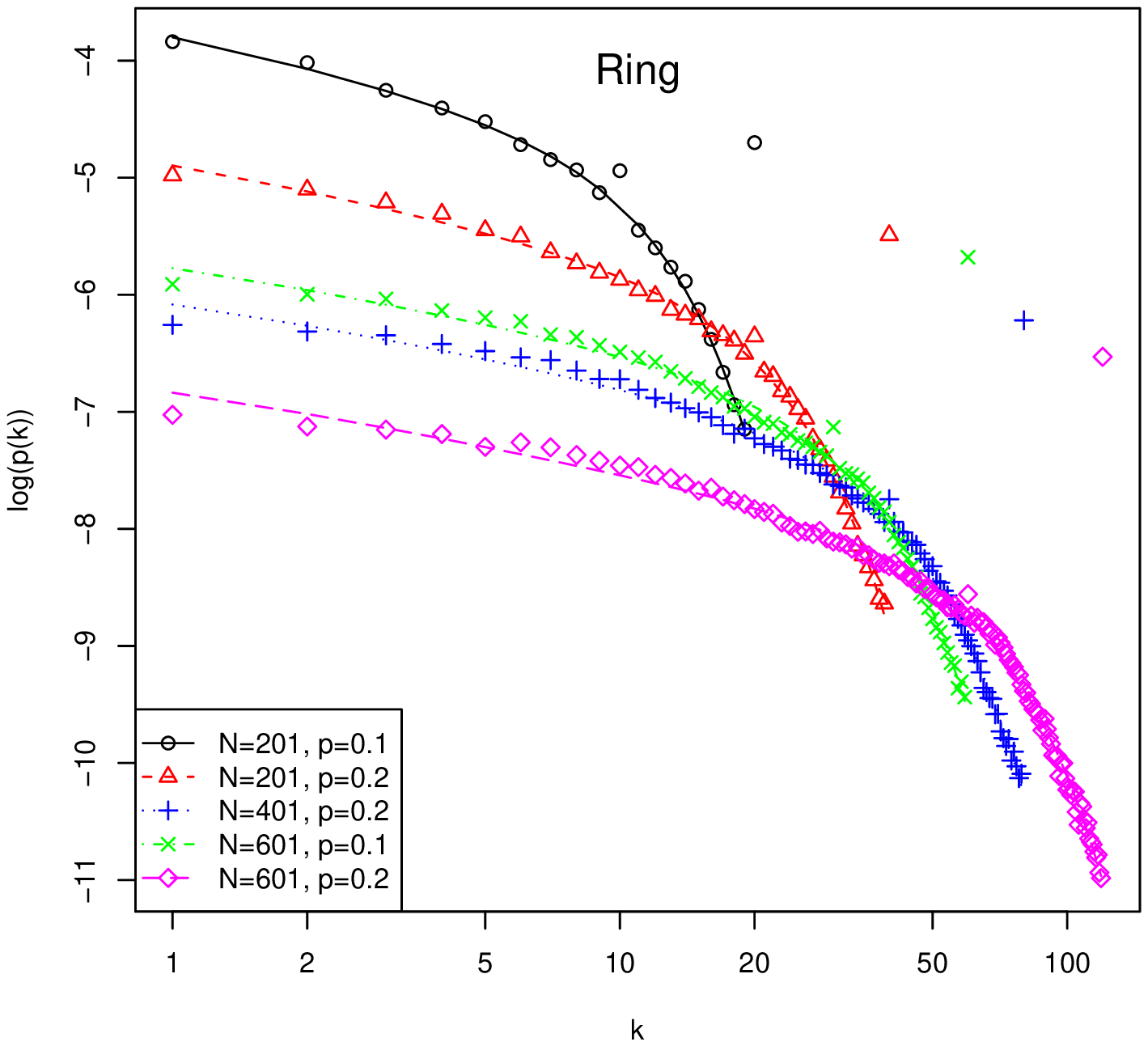}
                        \caption{Normalised influence network degree distributions for random (ER), scale-free (SF) and regular-ring (Ring) substrate networks. Each data set was generated for $m=6$, $S=2$ with a settling time of $100000$ time steps, time averaged over $1000$ influence networks separated by $100$ time steps each and ensemble averaged over $20$ realisations of the game. The standard error for each data point is not shown since it is smaller than the point itself in all cases.  The lines are the best fits using the forms given in Table~\ref{tab:fit-parameters}.}
                        \label{f:pk}
                    \end{center}
                \end{figure*}

                    Given the hypothesis that the local copying process seen in the Moran model explains the influence network structure for a random substrate network, each of the data sets shown in Figure~\ref{f:ERpkp1} was fitted to the function $p(k) \sim k^{-\gamma}e^{-\alpha k}$ of the Moran model \cite{E04,E07,EP07,EPY10}. Unfortunately, the fits were unsuccessful because of the tail. Through trial and error, the influence structure was found to better fit a function with a faster power law cutoff,
                    \begin{equation}
                        p(k) = A(k+k_{1})^{-\gamma}\left(1+\left(\frac{k}{k_2}\right)^\chi \right)
                        \label{eq:dpfit}
                    \end{equation}
                    where $A$, $k_{1}$, $k_{2}$, $\gamma$ and $\chi$ are parameters to be found. The resulting fit parameters for each of the random substrate network parameter sets, $(N, \ksav )$, are shown together with their associated errors in Table~\ref{tab:fit-parameters}.  Note that this confirms the suggestion made in \cite{ATBK04} that the initial fall off is power law with a power $\gamma \approx 1.0$. Figure~\ref{f:pk} shows the influence network degree probability distribution in log space with the functional fits superimposed.

                \subsection{Scale-Free Substrate Network}
                \label{ssub:scale_free_substrate_network}

To investigate the effect of having a much larger maximum degree and a broad degree distribution, the next substrate network investigated was a scale-free network. Figure~\ref{f:pk} shows the resulting influence structure for similar parameter sets, $(N, \ksav )$, as those used in the random network case.  Again the data was fitted to various forms but the best fit was also found with the double power law form of \tref{eq:dpfit} used for the random case above.  Table~\ref{tab:fit-parameters} gives the best fit parameters and associated errors. The initial scale-free leadership structure is still present, again with power $\gamma \approx 1$. However comparing the data for random and scale-free substrate cases, Figure~\ref{f:pk} shows the latter has a much sharper turnover and the fit is relatively poor for a few values at this transition.  Nevertheless the double power law form \tref{eq:dpfit} still provides a good description of the results.

                \subsection{Regular-Ring Substrate Network}
                \label{ssub:regular_ring_substrate_network}
                    Finally we look at a regular substrate network with much more local structure and no variation in degree compared to the previous two substrate types.   The influence structure for the regular-ring substrate network has isolated peaks at $\ksav $ and $\frac{\ksav }{2}$. If we ignore these extraneous points, the remaining data is similar in shape as before but now the turnover from power law to exponential decay is much more gradual than in the random and scale-free cases with no visually recognisable cutoff point. In this case we find that if we exclude the isolated peaks, the remaining points are best fit with an exponential cutoff
                    \begin{equation}
                        p(k) = A(k+k_{1})^{-\gamma}\exp \left\{ -(k/k_2)^\chi \right\}
                        \label{eq:pexpfit}
                    \end{equation}
                    with the best fit parameters and associated errors given in Table~\ref{tab:fit-parameters}.
                    Since $\chi$ is significantly different from $1.0$ the tail is still different from that in the Moran model, just as for the last two types of substrate network.  However while we still find the same scale-free leadership structure as before, it is now much less fat with $\gamma \approx 0.4$. Again, the resulting curves are superimposed on the data in Figure~\ref{f:pk}.

            \subsection{Data Collapse}
            \label{sub:data_collapse}

By looking at the fitted parameters it is clear that in all cases the $k_1$ parameter is marginal and can be dropped (set to zero) without any significant loss of precision.  The key features in all cases are an initial power law drop in $p(k)$, with power set by $\gamma$, followed by a sharp cutoff setting in at about $k_2$.

                \begin{table*}
                    \begin{center}
                        \begin{tabular}{|c|c|c|c||c|c|c|c|c|}
                                \hline
                                Type & $N$ & $\ksav $ & $p$ & $k_{s,\mathrm{min}}$ & $k_{s,\mathrm{max}}$ & $d$ & $\langle l\rangle$ & $\langle c \rangle$ \\ \hline
                                  & 201  & 20.1  & 0.10 & 9.3   & 33.7  & 3 & 2.0 & 0.10 \\ \cline{2-9}
                                  & 201  & 39.8  & 0.20 & 25.0  & 56.4  & 3 & 1.8 & 0.20 \\ \cline{2-9}
                              ER  & 601  & 60.1  & 0.10 & 36.7  & 85.5  & 3 & 1.9 & 0.10 \\ \cline{2-9}
                                  & 601  & 120.0 & 0.20 & 90.2  & 151.0 & 2 & 1.8 & 0.20 \\ \cline{2-9}
                                  & 1001 & 100.1 & 0.10 & 70.8  & 132.8 & 3 & 1.9 & 0.10 \\
                                \hline
                                  & 201  & 20.8  & 0.10 & 10.0    & 75.4  & 3    & 2.0 & 0.18 \\ \cline{2-9}
                                  & 201  & 42.3  & 0.21 & 21.1  & 114.3 & 3    & 1.8 & 0.31 \\ \cline{2-9}
                              SF  & 401  & 86.0  & 0.22 & 45.6  & 228.9 & 2.45 & 1.8 & 0.31 \\ \cline{2-9}
                                  & 601  & 62.4  & 0.10 & 32.5  & 240.9 & 3    & 1.9 & 0.18 \\ \cline{2-9}
                                  & 601  & 129.8 & 0.22 & 71.9  & 344.2 & 2    & 1.8 & 0.31 \\\cline{2-9}
                                  & 1001 & 104.0 & 0.10 & 54.9  & 388.4 & 3    & 1.9 & 0.18 \\
                                \hline
                                  & 201  & 20  & 0.1 & 20  & 20  & 10 & 5.5 & 0.71 \\ \cline{2-9}
                                  & 201  & 40  & 0.2 & 40  & 40  & 5  & 3.0 & 0.73 \\ \cline{2-9}
                            Ring  & 601  & 60  & 0.1 & 60  & 60  & 10 & 5.5 & 0.74 \\ \cline{2-9}
                                  & 601  & 120 & 0.2 & 120 & 120 & 5  & 3.0 & 0.74 \\ \cline{2-9}
                                  & 1001 & 100 & 0.1 & 100 & 100 & 10 & 5.5 & 0.74 \\
                                 \hline
                            \end{tabular}
                            \label{tab:regular-ring-network-properties}
                        \caption{Properties of the substrate network for the relevant parameter sets, $(N,\kav )$, for random (ER), scale-free (SF) and regular-ring (Ring) networks. Here $p=\ksav/(N-1)$, $k_{s,\mathrm{min}}$ and $k_{s,\mathrm{max}}$ are minimum and maximum degree, $d$ is the diameter,  $\langle l\rangle$ the average shortest path between vertex paths and $\langle c \rangle$ is the average clustering coefficient. In each case, the quantities were averaged over $20$ realisations of the corresponding graph.}
                        \label{tab:substrate-network-properties}
                    \end{center}
                \end{table*}

\begin{figure*}[tbh!]
                    \begin{center}
                        \includegraphics[width=0.47\textwidth]{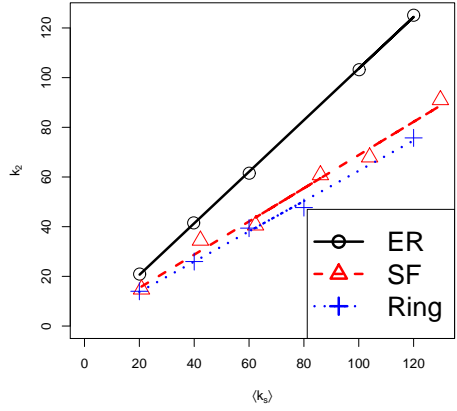}
                        \caption{Plots of the fit parameter, $k_{2}$, against average degree, $\kav $, for each of the random (black circles and solid line), scale-free (red triangles and dashed line) and regular-ring (blue plus and dotted line) substrate networks. The lines are the best linear fits. In each case, the standard errors are smaller than the data point and so are not shown.}
                        \label{fig:degree-correlation}
                    \end{center}
                \end{figure*}

Table~\ref{tab:substrate-network-properties} presents the main network characteristics for each of the substrate network types. Using this data it is possible to plot each of the fit parameters given in Table~\ref{tab:fit-parameters} against each of the network characteristics in an attempt to find a relationship.  The regular ring graph has a $\gamma$ which is markedly different from the other two substrate networks but otherwise $\gamma$ shows no strong dependence on the network parameters.  However there is a clear linear dependence of the cutoff scale $k_2$ on the average degree of the substrate network $\ksav$ as Figure~\ref{fig:degree-correlation} shows.  The coefficients do differ significantly between networks however and we find that
                \begin{eqnarray}
                    k_{2}^{\textrm{\tiny{random}}} &= 1.038(8) \ksav  - 0.2(6) \\
                    k_{2}^{\textrm{\tiny{scale-free}}} &= 0.67(4)\ksav  + 2(3) \\
                    k_{2}^{\textrm{\tiny{regular-ring}}} &= 0.61(2)\ksav  + 2(2) \, ,
                    \label{k2ksav}
                \end{eqnarray}
where the brackets give the standard errors in the last digit.  Note that a zero intercept is consistent here but what is important is that the cutoff scale $k_2$ is a linear function of $\ksav$ in each case.

This now suggests how we can summarise our results. The leadership distributions found can be well approximated by a power law section followed by essentially nothing,
\beq
 p(k) = \left\{\begin{array}{ccl}
  p(0) & \mbox{if} & k=0 \\
  Ak^{-\gamma}  & \mbox{if} & 0<k\leq k_2 \\
  0  & \mbox{if} & k> k_2 \\
  \end{array}\right. \, .
  \label{pkapprox}
\eeq
We have two constraints, that $\sum_k p(k) = 1$ and $\sum_k kp(k) = \kav \approx 1$ since for the influence network most vertices point to a single leader, while a few do not.  Approximating these sums as an integral gives us simple expressions for the two unknowns $A$ and $p(0)$
\beq
 A =\frac{\kav (2-\gamma)}{k_2^{2-\gamma}-1} \, , \qquad p(0) = 1- \frac{A}{1-\gamma} (k_2^{1-\gamma} -1) \, .
\eeq

In terms of the parameter $p=\ksav /(N-1)$, it was noted empirically in \cite{ATBK04} that $n(1) = Np(1) =  \frac{1}{c p}$ with $c$ a constant independent of $p$ (equivalently $\ksav$) and $N$.  Within our approximations we have that
\beq
 c \approx \frac{(k_2^{2-\gamma}-1)} {\ksav \kav (2-\gamma)} \frac{(N-1)}{N}
\eeq
For our examples, $k_2, N \gg 1$ are good approximations. For the
random and scale-free graphs (but not for the regular ring graphs)
the approximation $\gamma \approx 1$ is weaker but acceptable in
most cases.  This then leaves $c=(k_2/\ksav) (1/\kav)$.  The first
term is approximately constant for all networks though the value
does depend on the type of network as shown in \tref{k2ksav}. It
is not surprising that $\kav$ is close to one since in the
influence network each agent can follow at most one other agent.
So in our language, the result $p=\ksav /(N-1)$ of \cite{ATBK04}
follows if we assume the leadership distribution is a power law
followed by a sharp cutoff at $k_2$ which is proportional to the
average degree of the substrate network $\ksav$.  We have noted
exactly this behaviour for a wider range of networks in
Table~\ref{tab:fit-parameters} and
Fig.~\ref{fig:degree-correlation}.

\begin{figure*}[ptbh!]
                    \begin{center}
                        \includegraphics[width=0.49\textwidth]{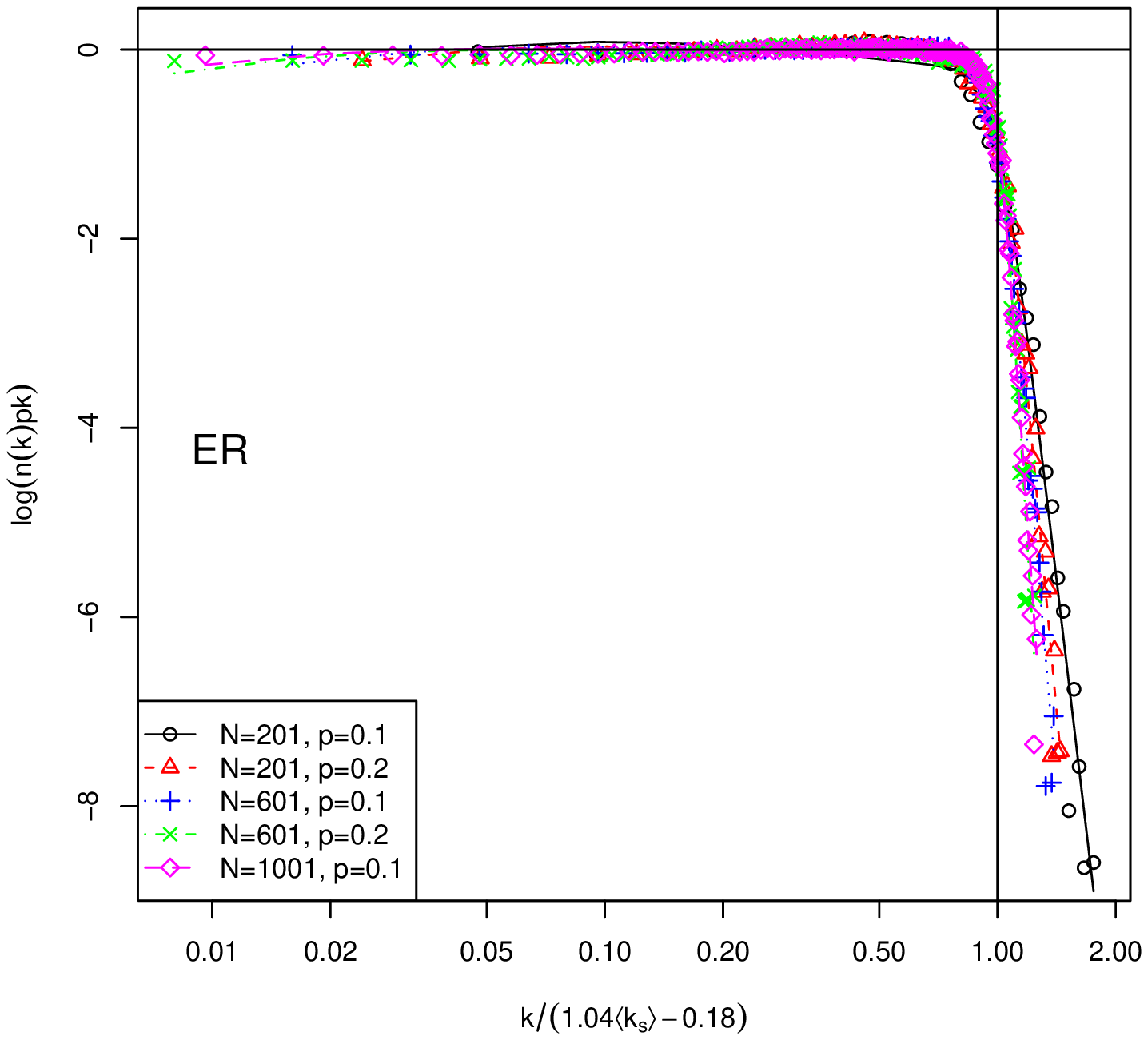}
                        \includegraphics[width=0.49\textwidth]{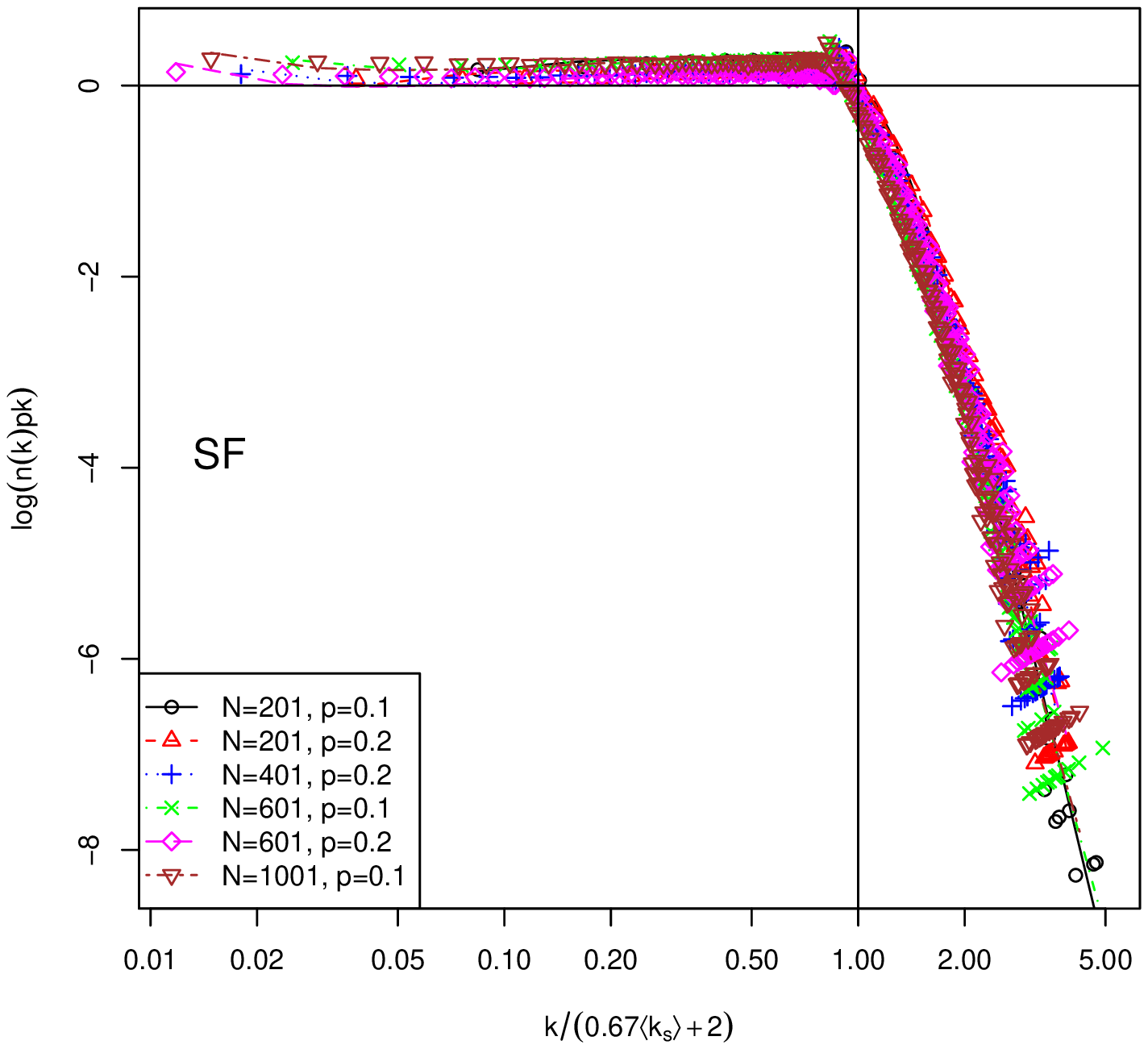}
                        \\
                        \includegraphics[width=0.49\textwidth]{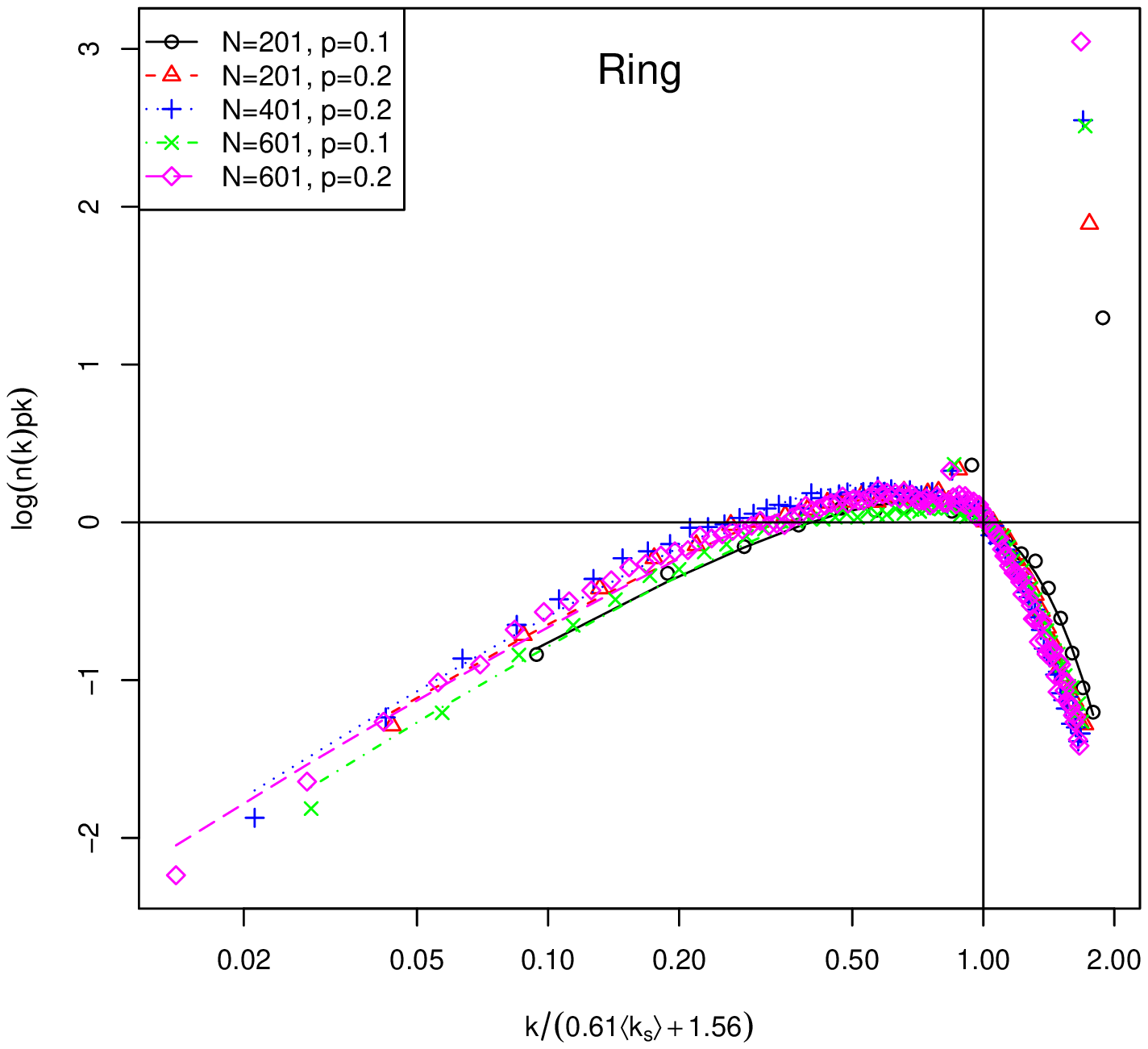}
                        \caption{Data collapse for the influence network degree distributions for for random (ER), scale-free (SF) and regular-ring (Ring) substrate networks.  The data sets are the same as used in plotting Figure~\ref{fig:degree-correlation}. Again, the standard error is smaller than each data point and thus is not shown.}
                        \label{f:pk-collapsed}
                    \end{center}
                \end{figure*}

In terms of data collapse it suggests that we plot $y= k p(k). N p$ against $x=k/k_2$ since \tref{pkapprox} then
becomes
\beq
 y(x) = (2-\gamma)\frac{N}{N-1} \frac{\ksav}{k_2}\frac{1}{1-k_2^{\gamma-2}} \; x^{1-\gamma}
 \, , \qquad
 0 < x \leq 1 \, .
 \label{collapsefunc}
\eeq
The coefficient should vary only with the type of substrate graph used, it should be approximately independent of $N$ and $\ksav$.
Replotting the data used in Figure~\ref{fig:degree-correlation} using the form \tref{collapsefunc} now gives the satisfying result shown in Figure~\ref{f:pk-collapsed}.

\begin{figure*}[tbh!]
                    \begin{center}
                        \includegraphics[width=0.49\textwidth]{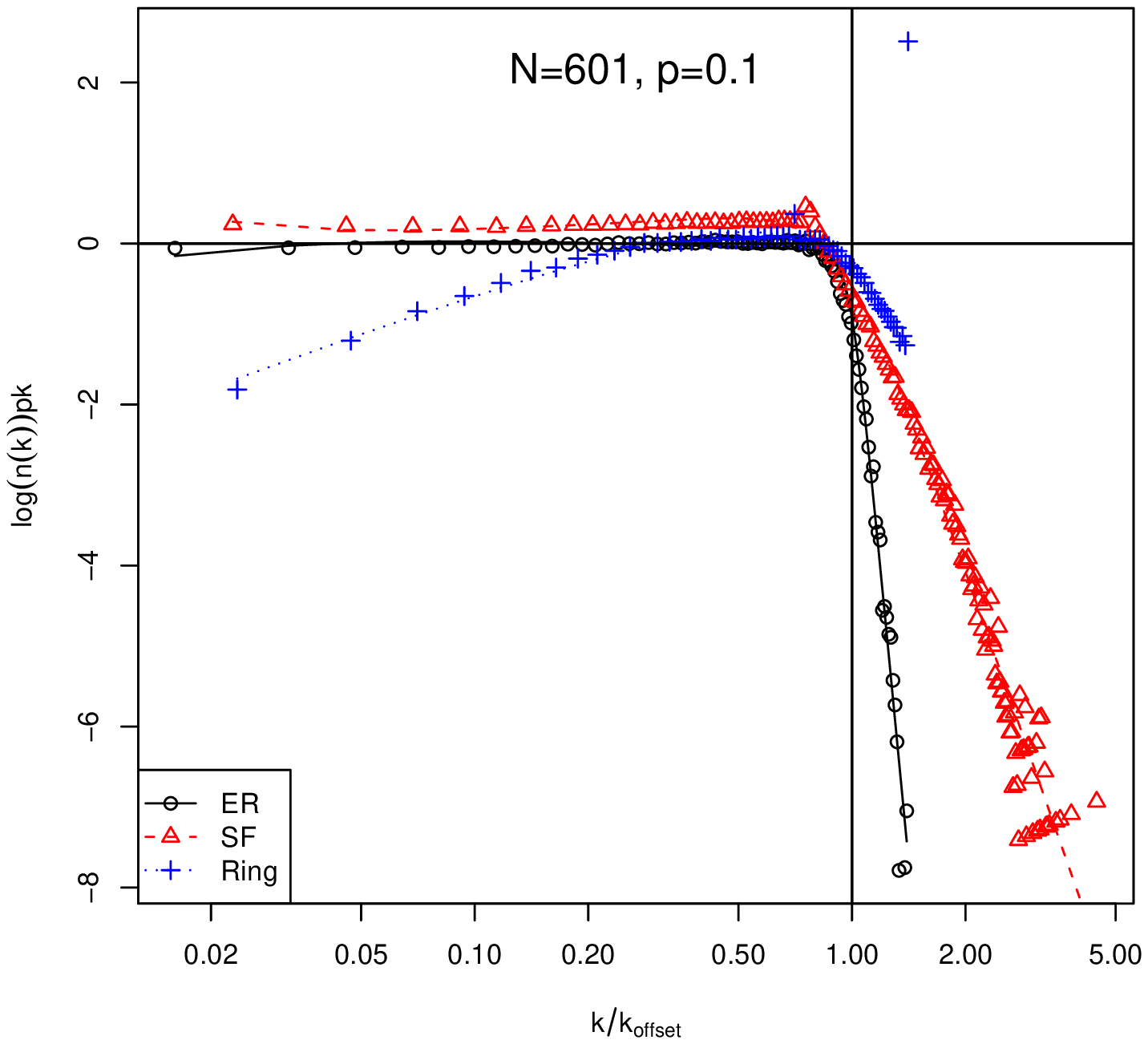}
                        \includegraphics[width=0.49\textwidth]{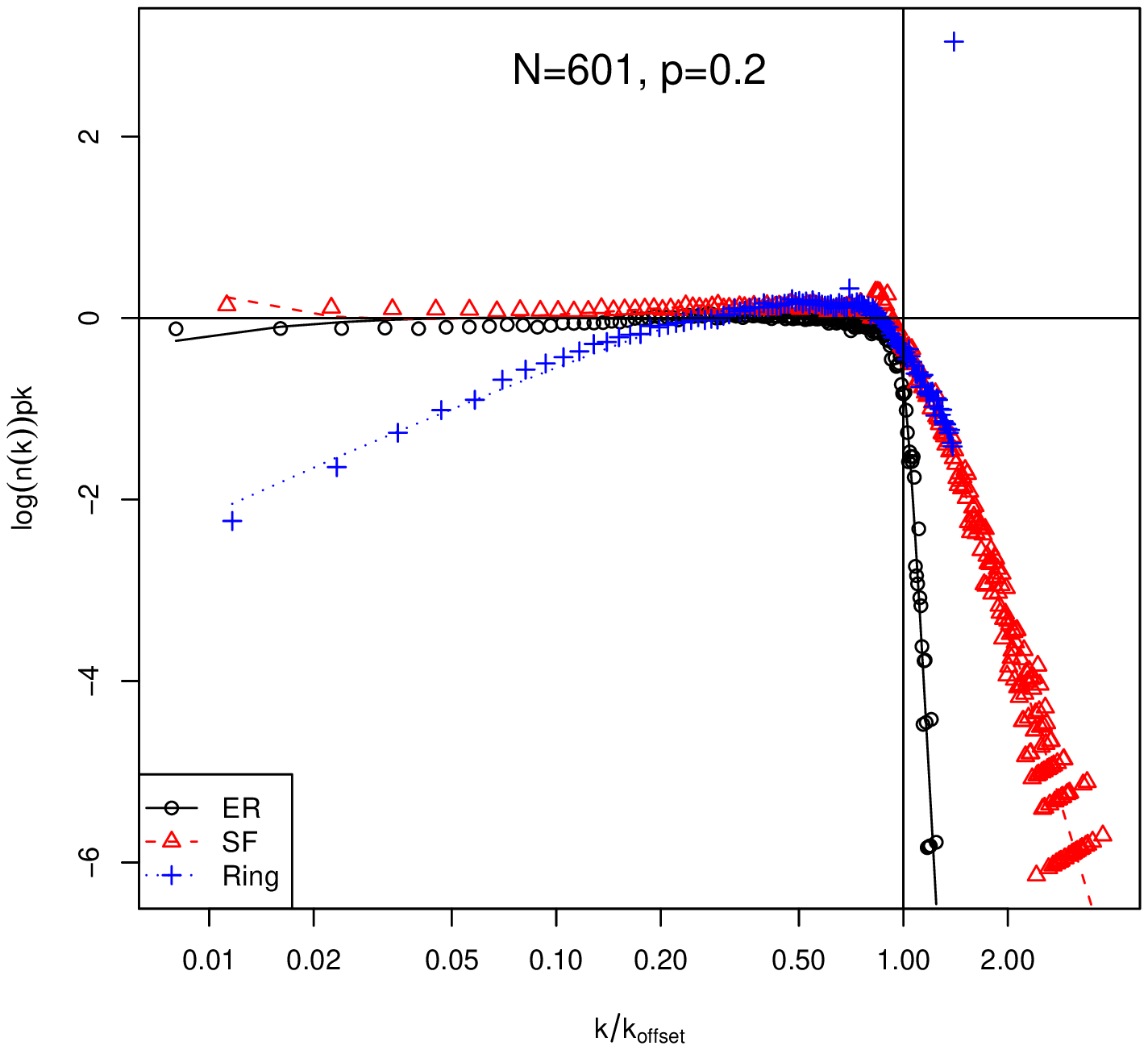}
                        \\
                        \includegraphics[width=0.49\textwidth]{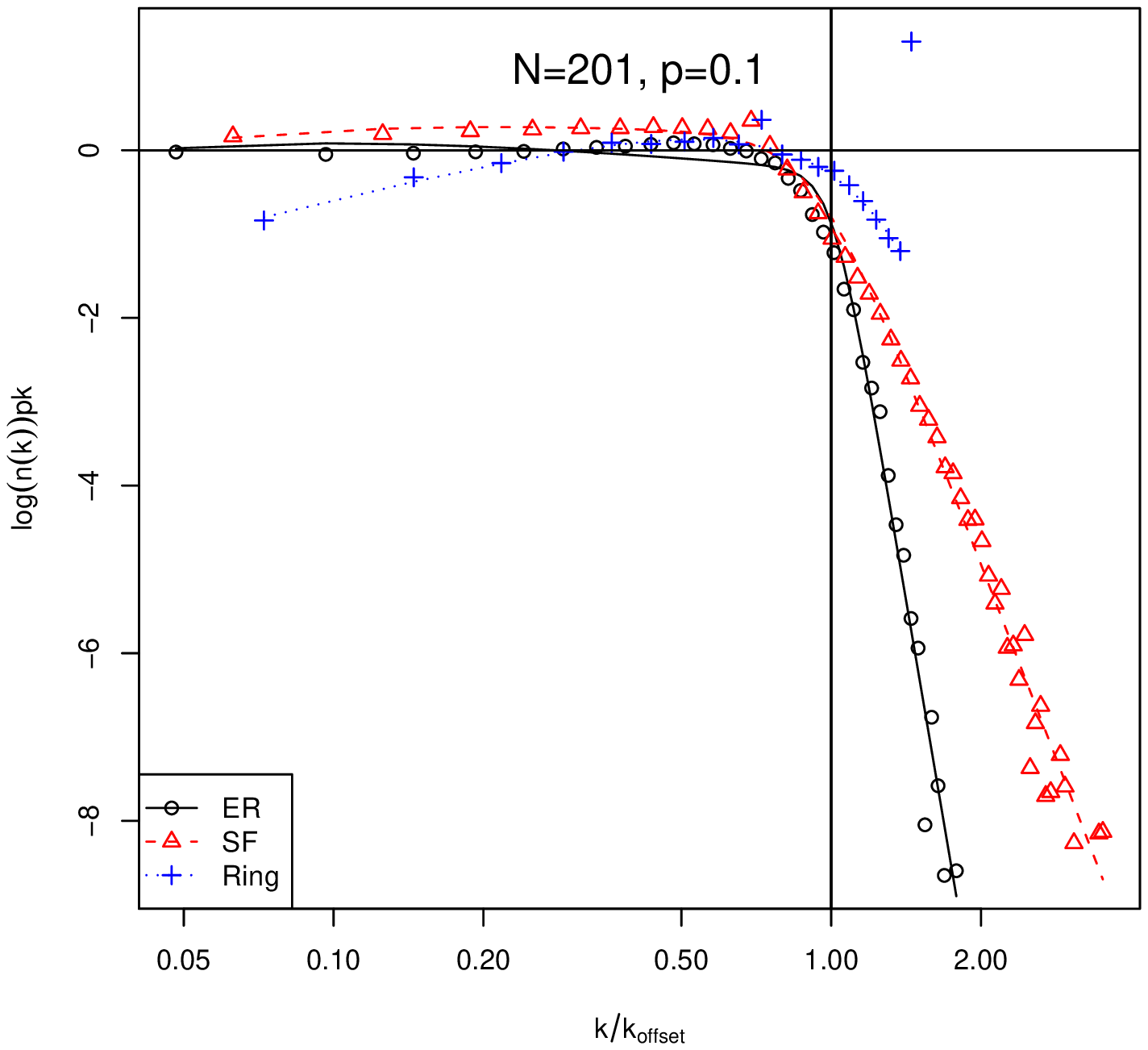}
                        \caption{Comparison of collapsed influence network degree distributions.  Each plot shows three different types of substrate network but each has the same number of vertices and approximately the same number of edges.
                        Top left has   $N=601$, $p\approx 0.1$, top right has  $N=601$, $p\approx 0.2$
                        and the bottom plot shows $N=201$, $p\approx 0.1$.
                        The data sets used are the corresponding sets for the random (ER), scale-free (SF) and regular-ring (Ring) substrate network as shown in Figure~\ref{fig:degree-correlation}.}
                        \label{fig:collapsed-comparison}
                    \end{center}
                \end{figure*}

What is clear is that there are relatively minor differences between the random graphs and the scale free graphs, but that there is a fundamental change when working with the regular ring graph.  To show this more clearly we plot the data for three different types of graph but with essentially the same number of vertices and edges using the data collapse function \tref{collapsefunc}.  The result is shown in Figure~\ref{fig:collapsed-comparison} and emphasises the differences between the different type of graph.

\section{Conclusions}\label{cha:conclusions}

We have studied the minority game in which the agents act on a
substrate network representing the social network of agents in the
market. We investigated three different types of substrate
networks: Erd\H{o}s-R\'{e}nyi random graph, scale-free and regular-ring, chosen to represent
various extremes in terms of their substrate degree distributions,
and investigated each using a variety of network sizes.  By
fitting the data to various functions we found two forms gave good
fits, \tref{eq:dpfit} and \tref{eq:pexpfit}, and that the
resulting degree distribution of the influence network may always
be characterised as having a power law distribution followed by a
sharp cutoff. This is best illustrated by replotting our data in
terms of the collapse function \tref{collapsefunc}, as shown in
Figures~\ref{f:pk-collapsed} and \ref{fig:collapsed-comparison}
which highlights the universal aspects of the behaviour.  In
particular our form \tref{collapsefunc} shows how the leadership
varies in a universal manner for one type of substrate network as
the number of vertices and edges (equivalently $N$ and $\ksav$)
changes.

The universal form shows that irrespective of the type of
substrate network, the leadership structure which emerges has an
initial power law section.  That is, there are always a few agents
whose actions are copied by many others.  In particular we find
this power law whether or not the underlying social network has a
scale-free structure, or if it is a regular graph, and thus have
significantly extended the finding of Anghel et al.\ \cite{ATBK04}
for the random substrate network. However we have also found that
the nature of the social network linking agents does have a
significant effect on the `fatness' of the leadership
distribution, that is the power $\gamma$ in our parameterisations
\tref{eq:dpfit}, \tref{eq:pexpfit}, \tref{pkapprox} or \tref{collapsefunc}.
In particular we find that
the regular graph does not show the power of $\gamma \approx 1$
found for the random and scale-free graphs here and noted for the
random graph in \cite{ATBK04}.

The manner in which this power law
in leadership is cutoff has also been quantified.  While the
nature of the tail depends on the type of substrate network, we
have shown that the cutoff point is always proportional to the
average degree of the underlying social network.

The minority game is clearly not realistic in many senses but it
can capture some of the basic features of competition dynamics.  In the
case of the networked minority game investigated here, it is clear
that the copying of strategies from a local social neighbourhood
does lead to the emergence of a leadership structure, regardless
of the nature of the social network.  On the other hand it appears
the copying processes represented by the Moran model
\cite{E04,E07,EP07,EPY10} are too simple to capture the full
detail of the leadership structure emerging in these networked
minority games.  Nevertheless the Moran model with its exact
analytic results does provide an explanation as to how a power law
in leadership emerges in the networked minority game.

We stress that the functional forms used to fit our results, \tref{eq:dpfit} and \tref{eq:pexpfit}, are statistically acceptable but are not the only possible forms which could be used.  We use them to provide a quantitative description of the leadership structure in the networked Minority game so that we can make quantitative comparisons for different types of networks and against the Moran model.  We do not derive any particular meaning from the particular functional forms used other than they describe leaderships structures with an initial power-law behaviour followed by a sharp cutoff \tref{pkapprox}.  This description is sufficient to give an excellent data collapse.  We have, though, excluded the simple Moran form as a good fit.

It would be interesting to change the way the information flows in the substrate network.  In particular both Cajueiro and De Camargo \cite{CD06} and Lavi\v{c}ka and Slanina \cite{LS07a} allow agents to follow the strategies of non-nearest neighbours. This is done by allowing agents to copy what we have termed the decisions of nearest neighbours.  Reality is likely to be somewhere in between, information can flow beyond nearest neighbours, but information on the decisions of most successful agent is unlikely to be transmitted perfectly over several steps in the substrate network. It would be interesting to revisit the different questions raised here and in \cite{ATBK04,CD06,LS07a} for these more realistic information flows.

\section{Acknowledgements}
    \label{cha:acknowledgements}

We would like to thank Imperial College Information and
Communication Technologies section for allowing us to use their
servers to perform the simulations. TC would also like to
acknowledge the online software development community for their
advice and guidance in the best way to construct the minority game
model to maximise its modularity, flexibility and overall
efficiency. The source code is open source under the MIT license
and is available to view online at
\texttt{http://www.github.com/tobyclemson/msci-project}.
\tprecomment{Further details are available in the Appendix below and in \cite{Clemson10}.}












%

\newpage
\renewcommand{\thesection}{A\arabic{section}}
\setcounter{section}{0}
\section*{Supplementary Material}

This supplementary material is not part of the published paper.

\section{Obtaining the Source Code} \label{cha:source_code}

All source code has been released open source under the MIT license and is available to view online at \texttt{http://www.github.com/tobyclemson/msci-project} via the github online source code repository. Alternatively, to download a complete working copy of the repository, the \texttt{git} version control tool is required, available from \texttt{http://git-scm.com/}.

        To clone the source code repository, execute the following at the command line of a computer with \texttt{git} installed:

        \begin{minipage}[c]{\textwidth}
            \hspace{0.8cm}\footnotesize{\texttt{> git clone \texttt{git://github.com/tobyclemson/msci-project.git} target\_directory}}
        \end{minipage}

        where \texttt{target\_directory} is the name of the directory you want the source code downloaded to.

        Once the repository has been cloned, a number of directories are available:

        \begin{minipage}[c]{\textwidth}
            \hspace{0.8cm}\footnotesize{\texttt{> tree -ad -L 1}}

            \hspace{0.8cm}\footnotesize{\texttt{  .}}

            \hspace{0.8cm}\footnotesize{\texttt{  |-- .git}}

            \hspace{0.8cm}\footnotesize{\texttt{  |-- features}}

            \hspace{0.8cm}\footnotesize{\texttt{  |-- lib}}

            \hspace{0.8cm}\footnotesize{\texttt{  |-- nbproject}}

            \hspace{0.8cm}\footnotesize{\texttt{  |-- spec}}

            \hspace{0.8cm}\footnotesize{\texttt{  |-- src}}

            \hspace{0.8cm}\footnotesize{\texttt{  \`{}-- vendor}}
        \end{minipage}

        The \texttt{.git} directory contains a complete copy of the source code repository including every version ever committed using the \texttt{git} version control system. A good introduction on using \texttt{git} can be found at \texttt{http://progit.org/book/}. The \texttt{src} directory contains all Java source files for the minority game model. The \texttt{spec} and \texttt{features} directories contain the unit and acceptance tests respectively, written using Ruby. The \texttt{vendor} directory contains all third party libraries used during development of the computational model. The \texttt{lib} directory contains some Ruby scripts that aided in testing the Java code base. Finally, the \texttt{nbproject} directory contains NetBeans specific files allowing the code to be imported into the NetBeans IDE.

For further details on the computational model, refer to \ref{cha:architecture_of_the_computational_model}.

        \section{Computational Model}
        \label{sec:computational_model}

            Here a very brief overview of the computational methods used in generating all data presented in Section~\ref{cha:results_and_observations} is given. The full source code is available online as detailed in \ref{cha:source_code}. For a more thorough technical coverage of the computational model, see \ref{cha:architecture_of_the_computational_model}. 

            \subsection{Implementation}
            \label{sub:implementation}

                Early on, it was decided that Java would be used as the main language for the computational model since it is easier and less error prone to develop with than C/C++. This ease of development comes at the cost of performance but it was felt that this would not be a problem since Java is already widely used in the scientific community \citep{BMPP01}.

                Java has numerous other benefits: it is portable allowing the code to be run on a wide variety of platforms, it has a very large base of existing libraries both standard to Java and developed by third parties and it is very widely used meaning that much help is available from online communities.

                The Ruby scripting language was also used in developing the computational model to aid in testing, prototyping and experimenting with the Java code base. This was made possible using a Ruby implementation called JRuby which allows the two languages to work together.

                To reduce the time spent developing the computational model, it was decided that custom code should only be written for the core minority game behaviour utilising existing libraries for everything else. The external libraries used were all very mature in their development and as a result, well tested and mostly bug free.

                The graph data structures and algorithms necessary for the networked minority game code were provided by the Java Universal Network/Graph Framework (JUNG) \cite{jung} which is full featured, fast and flexible. The library is capable of constructing and manipulating graphs of any sort. It also provides functionality to calculate various graph properties.

                The random number generator used was taken from the CERN Colt scientific library \cite{colt}.

                All statistical analysis was automated by the model in two ways. Firstly, many statistical quantities such as means, variances or standard errors were calculated inside the simulation using the classes available in the CERN Colt scientific library \cite{colt}. Secondly, when further processing was required on the generated data, it was exported to Excel spreadsheets complete with all required formulae. This was made possible by the Apache POI library.

                One major benefit of performing analysis inside the simulation was that the algorithms used could be programmatically tested reducing the possibility of human error.

                An incremental approach was used in the development, and so it was important to make the code as modular and extensible as possible and so a fully object oriented approach was adopted utilising a number of tried and tested design patterns. Full details of the design of the model are given in \cite{Clemson10}.

            \subsection{Testing and Verification}
            \label{sub:testing_and_verification}

The code base itself was tested using automated tests written in Ruby. Besides detecting bugs in the software, utilising automated tests helped to identify breakages as new functionality was added to the system. Thus, even though the tests took time to write, they saved much time in terms of debugging and development.

Whilst automated testing ensured the software worked as expected, it did not verify any assumptions made about the rules of the minority game. To verify the model constructed worked in the same way as those used by past researchers, it was necessary to generate some previously established results and compare the resulting plots. To do this, the graphs in Figures~\ref{fig:standard-minority-game-verification}~and~\ref{f:ERpkp1} were generated and compared and as is shown in Section~\ref{cha:results_and_observations}, this verified the model was indeed producing the same results.

                    Another aspect that caused problems was when choosing between strategies or friends that had identical scores. The algorithm employed was at first incorrect since it did not return each such entity with uniform probability if there were more than two with the same score. This problem was discovered using statistical unit testing and was solved by rewriting the algorithm.

            \subsection{Optimisation}
            \label{sub:optimisation}

As each component of the system was completed, preliminary simulation runs were executed in an attempt to gauge the time requirements for data generation. The preliminary runs for the networked game indicated that it would take in excess of 8 days to generate a data set comparable to that in the paper from \citet{ATBK04}. This execution time was despite running the simulation on a multicore server with ample memory.

The optimisation was performed by attaching a profiler to an example simulation run which could intercept and log method calls and requests for memory at well as perform temporal benchmarking. This made it possible to see which parts of the model were most CPU and memory intensive or took the longest time to run. Further, those parts could then be rewritten so that they performed in a more suitable time.

After all optimisations had been performed, the time taken to generate an influence network degree distribution had dropped to less than 4 days and further profiling indicated that the only remaining bottlenecks in the code were due to the third party libraries being used.

\subsection{Optimisation}
            \label{sub:verification}

\begin{figure*}[btph!]
                    \centering
                    \includegraphics[width=0.65\textwidth]{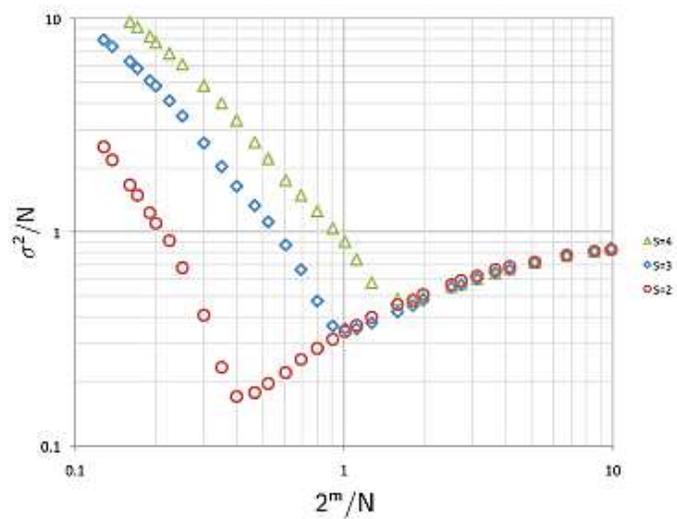} 
                    \caption{$\frac{\sigma^{2}}{N}$ for the standard minority game as a function of $S$, $N$ and $m$. The plot shows data points for $S=2,3$ and $4$ from bottom left to top right respectively. The key quantity here is $z=\frac{2^{m}}{N}$. As the plot shows, the minimum variance occurs for $S=2$, $z\approx{}0.4$. The simulation was run with a fixed memory capacity, $m=7$, allowing a settle time of $100\times{}2^{m} = 12800$ time steps. The variance for each $N$ was time averaged over $10000$ time steps and ensemble averaged over $100$ realisations of the game. The standard error in each data point is in all cases smaller than the data point itself.}
                    \label{fig:standard-minority-game-verification}
                \end{figure*}

For the standard minority game with no network, we studied the
normalised variance in choice attendance, $\frac{\sigma^{2}}{N}$,
as a function of memory capacity, $m$, number of agents, $N$, and
number of strategies per agent, $S$.
Figure~\ref{fig:standard-minority-game-verification} shows the
resulting plot and comparing to \citep{CMZ05}. As required, the
minimum variance occurs for $S=2$ and $\frac{2^{m}}{N} = 0.4$.
Note that this graph was generated for $m=7$ and the variances
were recorded after a settling time of $100\times2^{m}$ time steps
and ensemble averaged over $100$ realisations of the game.

                As can be seen from Figure~\ref{fig:standard-minority-game-verification},
                the optimum values of $S$, $m$ and $N$ are given by
                $S_{\textrm{\tiny{optimal}}}=2$ and $z_{\textrm{\tiny{optimal}}}=\frac{2^{m}}{N}=0.4$. Thus, having more than $2$ strategies does not benefit the agents as a collective. Similarly, the optimum memory capacity, $m_{optimal}$ is dependent on the number of agents in the game, $N$.

                Referring back to Section~\ref{sub:the_standard_minority_game}, it is clear the quantity $2^{m}$ is related to the size of the strategy space. In fact, it can be shown the strategies in a strategy space for a memory capacity $m$ are not all independent with the number of independent strategies being $2^{m}$ \citep{CZ98}. So the quantity, $\frac{2^{m}}{N}$, represents the ratio of independent strategies to the number of agents. For small $N$, only a fraction of the independent strategy space is spanned and the agents do not have enough collective intelligence to work together, instead acting more like random choice agents. As $N$ increases such that it is comparable to the number of independent strategies, a minimum is observed where the agents can cooperate efficiently. As $N$ continues to increase above the number of independent strategies, {herding} occurs wherein more than one agent has each independent strategy which is detrimental to the game \citep{CMZ05}.

\section{Architecture of the Computational Model}
    \label{cha:architecture_of_the_computational_model}

        This Appendix provides the more technical details of the design, implementation, testing and optimisation of the computational model used throughout the project.

        \subsection{Design}
        \label{sec:design}
            An object oriented design was used throughout. An inheritance hierarchy was built for the agent classes since each of the agents shared some common traits. Agent classes were implemented for random choice agents, basic agents that used only global information and networked agents that utilised the substrate network. The most important classes in the system are given in the following list:
            \begin{description}
                \item [\textsf{MinorityGame}:] The \texttt{MinorityGame} class is a controller for the minority game simulation. It provides methods to take a time step, determine the minority outcome and minority size as well as methods to access all other entities in the simulation such as agents, strategies, the substrate network etc.
                \item [\textsf{Agent}:] The \texttt{Agent} interface declares a number of methods that implementing classes must include to qualify as agents. The interface provides methods to ask an agent to prepare to make a decision, choose an outcome and update based on the minority choice as well as methods to handle agent identification, scores, predictions and decisions.
                \item [\textsf{BasicAgent}:] The \texttt{BasicAgent} class implemented the agent interface and made its decision for each time step using a brain only, i.e., it only made use of global information.
                \item [\textsf{NetworkedAgent}:] The \texttt{NetworkedAgent} class implemented the agent interface and made its decision using the predictions of the other agents with which it could communicate via the substrate network. The class also provides access to the agent's best friend at any point in time.
                \item [\textsf{Friendship}:] The \texttt{Friendship} class represents a friendship existing between two agents and is used in the social network to represent an edge.
                \item [\textsf{ChoiceMemory}:] The \texttt{ChoiceMemory} class implemented the memory functionality allowing past minority choices to be added and memory contents to be read. The memory worked like a fixed size pipe, the most recent minority choices are pushed in at the front and the oldest are pushed out and discarded from the back.
                \item [\textsf{Strategy}:] The \texttt{Strategy} class encapsulated the intelligence aspect of an agents brain as a mapping between possible memory contents and predictions.
                \item [\textsf{StrategyManager}:] The \texttt{StrategyManager} class manages an agent's strategies providing the ability to fetch the best performing strategy and update all strategies based on the past minority choice.
                \item [\textsf{Community}:] The \texttt{Community} class represents the entire community of agents along with the friendships between them containing the social network and providing access to everything inside the social network.
                \item [\textsf{Neighbourhood}:] The \texttt{Neighbourhood} class represents an agent's local neighbourhood of friends giving access to the best performing friend and all other local information available to the agents.
            \end{description}

By using classes to represent the memory, strategies and community, and local neighbourhood of agents it is very easy to switch in and out different functionality. This is an example of the \emph{Strategy} pattern. The \texttt{Agent} interface is actually a super-interface containing a number of smaller interfaces determining individual agent traits which can be implemented separately to give different types of agents. This is an example of the \emph{Facade} pattern. Abstract classes have also been used in a number of places implementing the core functionality of an interface so that subclasses don't have do implement quite so many methods. For further details on these design patterns see \citet{GHJV95}.

The minority game itself is composed of numerous entities such as agents, strategies, memories, scores, friendships etc. Each of the required simulations will use agents that choose in different ways or networks that are built in different ways etc. As such the construction of a minority game instance was abstracted away from the minority game itself into a number of factory classes that build the object hierarchy required by a single instance of the game. This allows many different game types to be created by simply changing the factories used. This is known as the \emph{Abstract Factory} pattern which decouples object behaviour from object creation.

In terms of the data collection aspect of the code it was hoped that there would be time to implement the \emph{Observer} pattern which allows different objects to register with another object to be notified when certain changes occur. In this way it would have been possible to attach data collectors to the minority game itself which would have been notified at the end of each time step and passed the minority game itself to update their local data stores. These data contained could then be passed to data analysers to perform all required analysis. Unfortunately there wasn't time to implement this so all data collection was performed inside the simulation harness that ran each simulation. This could be implemented in future.

            \subsection{Implementation}
            \label{sec:implementation}
                The computational model was implemented using Java.\footnote{Available from: \texttt{http://java.sun.com} [Accessed: 21st April 2010].} The advantages and disadvantages of Java are weighed up in the following listing:
                \subsubsection{Advantages}
                \label{ssub:advantages}
                    \begin{itemize}
                        \item Java avoids many of the low level aspects of languages such as C and C++ such as manual memory management and pointer based memory addressing, making it very easy and fast to develop with.
                        \item Java also has a very complete standard library of classes as well as many high quality third party open source libraries.
                        \item Java has many users worldwide so there is a good support network.
                        \item Specific to this project, there are a lot of graph theory related libraries available for Java.
                    \end{itemize}
                \subsubsection{Disadvantages}
                \label{ssub:disadvantages}

                    \begin{itemize}
                        \item Java is interpreted to an extent; the source code is compiled to an intermediate stage called bytecode which is interpreted on a virtual machine, thus making it slightly less efficient than C/C++.
                    \end{itemize}

A number of external libraries were used to speed up development. These are mature projects that are very well tried and tested. The following list details which aspects of the system they were used for:

                \begin{description}
                    \item [\sffamily Java Universal Network/Graph Framework (JUNG)] Rather than implement custom graph classes to represent the networks required, it made sense to use an existing library which was feature complete and well tested. The JUNG library offers a collection of interfaces as well as default implementations of a wide variety of graph data structures as well as algorithms for constructing and operating on those graphs.\footnote{Available from: \texttt{http://jung.sourceforge.net/} [Accessed: 21st April 2010].} The default implementations were more than suitable for the purposes of the minority game and by using this package much development time was saved.
                    \item [\sffamily CERN Colt] A number of issues were encountered with the random number generators included with the Java standard library. As a result, the more complete random number generation classes from the Colt library from CERN were employed which proved far more consistent than the built-in alternatives.\footnote{Available from: \texttt{http://acs.lbl.gov/~hoschek/colt/} [Accessed: 21st April 2010].}

                    The Colt library also contains a number of very useful statistical bin classes that automatically calculate a number of different statistical quantities for the data that they store. This allowed much of the data processing to occur inside the simulation itself.
                    \item [\sffamily Apache POI] The Apache POI library provides an Application Programming Interface (API) allowing Microsoft Excel documents to be created programmatically.\footnote{Available from: \texttt{http://poi.apache.org/} [Accessed: 21st April 2010].} Using this API it was possible to further automate the required data processing by automatically generating Excel spreadsheets containing all of the data for a particular run, complete with formulae to calculate errors and averages. After performing a run of the model, all that remained was for plots to be created manually inside the excel spreadsheet and for functional fits to be made to the resulting degree distributions as described in Section~\ref{cha:results_and_observations}.
                \end{description}

            \subsection{Testing}
            \label{sec:testing}
                All testing was performed using the Ruby scripting language.\footnote{Available from: \texttt{http://www.ruby-lang.org/en/} [Accessed: 21st April 2010].} This was realised by using a Ruby\footnote{Available from: \texttt{http://www.ruby-lang.org/en/} [Accessed: 21st April 2010].} implementation called JRuby\footnote{Available from: \texttt{http://jruby.org} [Accessed: 21st April 2010].} which allows Java and Ruby to operate alongside each other and have access to each others functionality.

When developing the code base, \emph{test driven development} was used as much as possible. This is a development practice where automated tests are written before the code under test. In this way, the tests act like a specification for the codebase and writing what you want the code to do before you write the code helps you to design the resulting code from the point of view of a client of that code.

The testing libraries that facilitated this style of development are RSpec\footnote{Available from: \texttt{http://rspec.info/} [Accessed: 22nd April 2010].} and Cucumber\footnote{Available from: \texttt{http://cukes.info/} [Accessed: 22nd April 2010]}, which test different aspects of the software.

RSpec tests the code at \emph{unit} level which corresponds to one set of tests per class. In this way, the behaviour of the class can be verified in isolation from the rest of the code base. A number of different types of unit tests were used:
                \begin{description}
                    \item [\sffamily State based:] ensuring that given a particular input, a method on an object of the class in a particular state produces the correct output.
                    \item [\sffamily Interaction based:] ensuring that a method on a class calls a method on one of its dependencies.
                    \item [\sffamily Performance based:] ensuring a method executes within some time frame.
                    \item [\sffamily Statistics based:] ensuring that probabilistic algorithms performed correctly based on the statistics of random variables.
                \end{description}

In contrast, Cucumber allows \emph{acceptance} tests to be written which verify that all classes work together correctly at the level of the entire model. These functional tests ensure that the distinct entities in the system produce the desired results at the scale of the entire system.

\subsection{Optimisation}
            \label{sec:optimisation}
                All optimisation was performed using the NetBeans IDE.\footnote{Available from: \texttt{http://netbeans.org/} [Accessed: 22nd April 2010].} This allow the code to be profiled by attaching a profiler that gives detailed and real time benchmarking data on the number of method calls made to particular methods, the CPU time occupied by each part of the system, the memory consumption and the time taken by each part of the simulation. This in turn allowed the code to be refactored by optimising at the method level cutting execution time by over 50\% in many cases.

\end{document}